\documentclass[a4paper,11pt]{article}
\pdfoutput=1 
\usepackage{jcappub} 
\usepackage{longtable}
\usepackage{float}
\usepackage{dcolumn}
\usepackage{bm}
\usepackage{appendix}
\usepackage{multirow}
\usepackage{color}
\usepackage{soul}
\usepackage[normalem]{ulem}
\usepackage[multiple]{footmisc}
\usepackage[utf8]{inputenc}
\usepackage{footmisc}

\newcommand{\Mpc}{\mathrm{Mpc}}

\title{The $H_0$ and $\sigma_8$ tensions and  the scale invariant spectrum}

\author[a]{M. Benetti} \emailAdd{micolbenetti@on.br}
\author[a,b]{Leila L. Graef} \emailAdd{leilagraef@on.br}
\author[a,c,d]{J. S. Alcaniz}\emailAdd{alcaniz@on.br}

\affiliation[a]{Departamento de Astronomia, Observat\'orio Nacional, 20921-400, Rio de Janeiro, RJ, Brasil}
\affiliation[b]{Departamento de F\'{\i}sica Te\'orica, Universidade do Estado do Rio de Janeiro, 20550-013 Rio de Janeiro, RJ, Brasil}
\affiliation[c]{Physics Department, McGill University, Montreal, QC, H3A 2T8, Canada}
\affiliation[d]{Departamento de F\'\i sica, Universidade Federal do Rio Grande do Norte, 59072-970 Natal, RN, Brasil}

\abstract{

In a previous communication~\cite{Benetti:2017gvm} it was shown that a joint analysis of Cosmic Microwave Background (CMB) data and the current measurement of the local expansion rate favours a model with a scale invariant spectrum (HZP) over the minimal $\Lambda$CDM scenario provided that the effective number of relativistic degrees of freedom, $N_{eff}$, is taken as a free parameter. Such a result is basically obtained due to the Hubble Space Telescope (HST) value of the Hubble constant, $H_0 = 73.24 \pm 1.74$ $\rm{km.s^{-1}.Mpc^{-1}}$ (68\% C.L.), as the CMB data alone discard the HZP+$N_{eff}$ model. Although such a model is not physically motivated by current scenarios of the early universe, observations pointing to a scale invariant spectrum may indicate that the origin of cosmic perturbations lies in an unknown physical process. Here, we extend the previous results performing a Bayesian analysis using joint CMB, HST, and Baryon Acoustic Oscillations (BAO) measurements. In order to  take into account the well-known  tension on the value of the fluctuation amplitude parameter, $\sigma_8$, we also consider Cluster Number counts (CN) and Weak Lensing (WL) data.  We use two different samples of BAO data, which are obtained using two-point spatial (BAO 2PCF) and angular (BAO 2PACF) correlation functions. Our results show that the joint CMB+HST+WL+NC dataset favor the extensions of the $\Lambda$CDM model over its minimal parameterization. Also, analysis with the BAO 2PCF always discard the HZP$+N_{eff}$ model with respect to standard scenario, whereas the combinations using BAO 2PACF favor the former model. 
We, therefore, find that all dataset  disfavor the $\Lambda$CDM model with respect to the  HZP$+N_{eff}$ extension, the only exception being the joint analysis with  BAO (2PCF).
}

\begin{document}
\maketitle
\section{Introduction}

Recently, several works  have focused on the so-called $H_0$ tension, the discrepancy on the value of the Hubble constant, $H_0$, obtained 
in the context of the $\Lambda$CDM model from current Cosmic Microwave Background (CMB) data~{\cite{Ade:2015xua},
$H_0 = 67.31 \pm 0.96$ $\rm{km~s^{-1}~Mpc^{-1}}$  (68\% C.L.), and the value measured using different geometric distance calibrations of Cepheids, $H_0 = 73.24 \pm 1.74$ $\rm{km~s^{-1}~Mpc^{-1}}$ (68\% C.L.)~\cite{Riess} (see also \cite{Riess:2018byc}). Currently, it is not completely clear whether such a tension is due to systematics in the {{data}} or an indication of new physics to be discovered beyond the standard model of cosmology (see, e.g.,~\cite{Bonvin:2016crt, Hildebrandt:2016iqg, Joudaki:2016mvz, Alsing:2016hkh, Kohlinger:2017sxk, Alam:2016hwk,Bernal:2016gxb, DiValentino:2016hlg, Tram:2016rcw, DiValentino:2017iww, Buen-Abad:2017gxg}).

In a previous communication~\cite{Benetti:2017gvm} we {{analysed a model proposed by Harrison \cite{Harrison:1969fb}, Zeldovich~\cite{Zeldovich:1972ij} and  Peebles~\cite{Peebles:1970ag} (hereafter HZP) before the birth of the
cosmological inflation theory. As well known, in the latter the scalar perturbations lead to a scale dependent primordial spectrum, so the scale invariant assumption, $n_{s}=1$, of the HZP model is
not supported within the inflationary context being necessary some unknown
fundamental process to justify such a spectrum. The current CMB data provide
a significant detection of a deviation of the scalar spectral index, $n_{s}$, from the
exact scale invariance value ~\cite{Ade:2015xua}, ruling out the HZP model. 

On the other hand,
the tension between the CMB and the local measurements of the Hubble constant
gave rise to investigations concerning the positive correlation between the
$H_{0}$ parameter with the spectral index ~\cite{Gerbino:2016sgw}, and we have seen that it can lead to a reconsideration of the HZP model which, if confirmed, would have a huge impact in the current cosmology. }}
In Ref.~\cite{Benetti:2017gvm} we pointed out that the scale invariance assumption, $n_s =1$, although not physically motivated by current scenarios of the early universe, is able to alleviate the $H_0$ tension when the model is extended with the  effective number of relativistic degrees of freedom, $N_{eff}$, hereafter HZP+$N_{eff}$ model, being $N_{eff} \simeq 3.7$ {\footnote{We have also shown that other simple extensions of the standard cosmology are not able to provide the same result. For instance, the HZP+$Y_{P}$ model, where $Y_P$ is the primordial helium mass fraction, is able to reconcile the local value of $H_0$ with CMB data only for values of $Y_P$ outside the Big Bang Nucleosynthesis allowed interval.}. 
Indeed, the $N_{eff}$ parameter is almost degenerated with $n_s$, altering the damping tail of the temperature spectrum and mimicking the role of a spectral index~\cite{Ade:2015xua, Verde:2013cqa, Hou:2011ec, Trotta:2003xg}, being also correlated with the $H_0$ value, since a higher $N_{eff}$ in the early universe leads to a faster expansion rate. 

Values of $N_{eff}> 3$ are allowed by recent Big Bang Nucleosynthesis (BBN) and CMB data~\cite{Ade:2015xua, DiValentino:2016hlg, Nollett:2013pwa,Benetti:2013wla, Hinshaw:2012aka, Boehm:2012gr, Cheng:2013csa}. 
{{From the theoretical point of view, standard particle physics predicts  a contribution  $N_{eff}=3.046$ due only to the three family of neutrinos that were relativistic at decoupling epoch. Nevertheless, even within the context of standard particle physics there can be extra contributions to the effective number of relativistic species coming from, for example, the stochastic background of primordial gravitational waves, which in some models can give a significant contribution to the value of $N_{eff}$~\cite{cabass, meerburg}. There is also the possibility of contributions to $N_{eff}$ which are not predicted by standard particle physics like, for example, eV-scale sterile neutrinos, thermal axions, Goldstone Bosons and even relativistic dark matter. These additional free-streaming relativistic particles are often called dark radiation~\cite{venninDR}.}}

\begin{table}[!t]
    \centering
    \begin{tabular}{|c|c|}
        \hline
        Parameter   & Prior \\
        \hline  
        \hline
		$100\,\Omega_b h^2$ 	
		&[ $ 0.005 : 0.1 $ ]
		\\
		$\Omega_{c} h^2$	
		&[ $ 0.001 : 0.99 $ ]
		\\
		$100\, \theta$ 
		&[ $ 0.5 : 10 $ ]
		\\
		$\tau$
		&[ $ 0.01 : 0.8 $ ]
		\\
		$\ln 10^{10}A_s$
		&[ $ 2.0 : 4.0 $ ]
		\\
		$N_{eff}$ 
		&[ $ 2 : 5 $ ]\\
		\hline  
    \end{tabular}
    \caption{\label{tab:priors} Priors on the model parameters}
\end{table}

Another ongoing issue in the $\Lambda$CDM context concerns the current large-scale structure data, more specifically to the value of the matter fluctuation amplitude on scales of $8h^{-1}\rm{Mpc}$, $\sigma_{8}$.   This issue is also related with the $H_0$ tension mentioned above, as the Hubble parameter is correlated with the matter density $\Omega_{m}$ and $\sigma_{8}$.  
Constraints on the $\Omega_{m}$-$\sigma_{8}$ plane have been widely discussed in the literature~\cite{salvati, MacCrann:2014wfa, Dossett:2015nda, Hamann:2013iba, Battye:2013xqa,Battye:2014qga} since the Planck primary CMB data provided significantly different constraints from the thermal Sunyaev-Zel'dovich cluster counts results~\cite{Ade:2015fva, Ade:2013lmv, Aghanim:2015eva} and the galaxy weak lensing~\cite{Heymans:2012gg}, these two latter preferring lower values of $\sigma_{8}$.  This tension is of particular interest since  galaxy clusters and weak lensing observations are  sensitive  to  the  cosmic matter density and $\sigma_{8}$ and offer a powerful complementary probe to CMB data and geometric observables. 

In this paper, we extend and complement the results of \cite{Benetti:2017gvm} considering both the $H_0$ and $\sigma_8$ issues above discussed. We perform a Bayesian analysis to test the observational viability of a scale invariant spectrum when the effective number of relativistic degrees of freedom is taken as a free parameter. We use current CMB data from the Planck collaboration~\cite{Aghanim:2015xee}, two sets of measurements of the Baryon acoustic oscillations (BAO)~\cite{Beutler:2011hx,Ross:2014qpa,Anderson:2013zyy,gabimicol, gabi, newgabi}, weak lensing~\cite{Heymans:2012gg} and Sunyaev-Zeldovich cluster counts~\cite{Ade:2015fva} observations and the latest value of the Hubble constant obtained by the Hubble Space Telescope (HST)~\cite{Riess}. It is worth mentioning that the two BAO data sets differ also in methodology. While the measurements of Ref.~\cite{Beutler:2011hx,Ross:2014qpa,Anderson:2013zyy} are obtained from a 3-dimensional analysis which calculates the two-point spatial correlation function (2PCF) and adopts the standard cosmology as fiducial model  -- to convert  the measured angular positions and redshift into comoving distances (hereafter 2PCF BAO), the BAO measurements of Ref.~\cite{gabimicol, gabi, newgabi} are obtained from a 2-dimensional analysis which calculates the two-point angular spatial correlation function (2PACF) without adopting any fiducial cosmology (we refer the reader to \cite{gabimicol, Salazar-Albornoz:2016psd} for a discussion on this subject).

\begin{table*}[!]
\centering
\caption{{
$68\%$ confidence limits for the cosmological parameters  using PLC+HST(+BAO) data. 
The $\Delta \chi^2_{best}$ and the $\ln {B}_{ij}$ refer to the difference of the HZP+$N_{eff}$ using BAO 2PCF (BAO 2PACF) with respect to the $\Lambda$CDM using BAO 2PCF (BAO 2PACF).}
\label{tab:Tabel_PLC+HST(+BAO)}}
\scalebox{0.7}{
\begin{tabular}{|c|c|c|c|c|}
\hline
\multicolumn{5}{|c|}{base dataset: PLC+HST}\\
\hline
{Parameter}&
{\textbf{$\Lambda$CDM (BAO 2PCF)}}&
{\textbf{$\Lambda$CDM (BAO 2PACF)}}& 
{\textbf{HZP+$N_{eff}$ (BAO 2PCF)}}&
{\textbf{HZP+$N_{eff}$ (BAO 2PACF)}}
\\
\hline
$100\,\Omega_b h^2$ 	
& $2.233 \pm 0.020$
& $2.277 \pm 0.020$ %
& $2.285 \pm 0.019$ %
& $2.301 \pm 0.019$ %
\\
$\Omega_{c} h^2$	
& $0.1183 \pm 0.0012$ %
& $0.1120 \pm 0.0013$ %
& $0.1300 \pm 0.0030$ 
& $0.1209 \pm 0.0028$ 
\\
$100\, \theta$ 
& $1.04106 \pm 0.00041$
& $1.04190 \pm 0.00041$ 
& $1.03998 \pm 0.00048$ 
& $1.04097 \pm 0.00048$ 
\\
$\tau$
& $0.083 \pm 0.017$
& $0.112 \pm 0.019$ 
& $0.102 \pm 0.018$
& $0.114 \pm 0.018$
\\
$\ln 10^{10}A_s$  \footnotemark[1]
\footnotetext[1]{$k_0 = 0.05\,\Mpc^{-1}$.}
& $3.096 \pm 0.035$ 
& $3.141 \pm 0.037$ 
& $3.161 \pm 0.035$ 
& $3.167 \pm 0.034$ 
\\
$n_{s}$
& $ 0.9690 \pm 0.0043$ 
& $ 0.9854 \pm 0.0048$ 
& $-$ 
& $-$ 
\\
$N_{eff}$
& fixed to $3.046$
& fixed to $3.046$ 
& $ 3.84 \pm 0.13$ 
& $ 3.58 \pm 0.12$ 
\\
$H_0$
& $ 67.97 \pm 0.55 $ 
& $ 70.94 \pm 0.63$ 
& $ 73.10 \pm 0.59 $
& $ 73.90 \pm 0.56 $
\\
\hline
\hline
$\Delta \chi^2_{\rm best}$         
& $ - $	    
& $-$ 	
& $ -6.2$   
& $+10.7 $   

\\
$\ln \mathit{B}_{ij}$ 
& $-$
& $-$  
& $ - 7.6$ 
& $+ 6.3 $   
\\
\hline
\end{tabular}}
\end{table*} 


\begin{table*}[!]
\centering
\caption{{
$68\%$ confidence limits for the cosmological parameters  using PLC+HST+WL+SZ(+BAO) data. 
The $\Delta \chi^2_{best}$ and the $\ln {B}_{ij}$ refer to the difference of the HZP+$N_{eff}$ using BAO 2PCF (BAO 2PACF) with respect to the $\Lambda$CDM using BAO 2PCF (BAO 2PACF).}
\label{tab:Tabel_PLC+HST+WL+SZ(+BAO)}}
\scalebox{0.7}{
\begin{tabular}{|c|c|c|c|c|}
\hline
\multicolumn{5}{|c|}{base dataset: PLC+HST+WL+SZ}\\
\hline
{Parameter}&
{\textbf{$\Lambda$CDM (BAO 2PCF)}}&
{\textbf{$\Lambda$CDM (BAO 2PACF)}}& 
{\textbf{HZP+$N_{eff}$ (BAO 2PCF)}}&
{\textbf{HZP+$N_{eff}$ (BAO 2PACF)}}
\\
\hline
$100\,\Omega_b h^2$ 	
& $2.240 \pm 0.019$ %
& $2.277 \pm 0.021$ %
& $2.263 \pm 0.020$ %
& $2.300 \pm 0.019$ %
\\
$\Omega_{c} h^2$	
& $0.1177 \pm 0.0013$ %
& $0.1119 \pm 0.0013$ %
& $0.1297 \pm 0.0027$ 
& $0.1215 \pm 0.0028$ 
\\
$100\, \theta$ 
& $1.04119 \pm 0.00037$
& $1.04190 \pm 0.00043$ 
& $1.03993 \pm 0.00052$ 
& $1.04079 \pm 0.00050$ 
\\
$\tau$
& $0.082 \pm 0.019$
& $0.106 \pm 0.019$ 
& $0.090 \pm 0.016$
& $0.107 \pm 0.021$
\\
$\ln 10^{10}A_s$  \footnotemark[1]
\footnotetext[1]{$k_0 = 0.05\,\Mpc^{-1}$.}
& $3.092 \pm 0.038$ 
& $3.128 \pm 0.038$ 
& $3.137 \pm 0.032$ 
& $3.153 \pm 0.041$ 
\\
$n_{s}$
& $ 0.9699 \pm 0.0042$ 
& $ 0.9855 \pm 0.0049$ 
& $-$ 
& $-$ 
\\
$N_{eff}$
& fixed to $3.046$
& fixed to $3.046$ 
& $ 3.84 \pm 0.13$ 
& $ 3.60 \pm 0.13$ 
\\
$H_0$
& $ 68.31 \pm 0.60 $ 
& $ 71.01 \pm 0.62 $ 
& $ 72.92 \pm 0.60 $ 
& $ 73.80 \pm 0.71 $
\\
$\sigma_{8}$
& $ 0.824 \pm 0.015$
& $ 0.821 \pm 0.015$  
& $ 0.869 \pm 0.015$ 
& $ 0.853 \pm 0.018$ 
\\
\hline
\hline
$\Delta \chi^2_{\rm best}$         
& $ - $	%
& $ - $ 
& $ - 13.4$
& $ + 0.3$   

\\
$\ln \mathit{B}_{ij}$ 
& $-$  %
& $-$  %
& $ - 3.9 $ %
& $ + 5.1 $ %
\\
\hline
\end{tabular}}
\end{table*} 
%

\begin{table}[!]
\centering
\caption{{
$68\%$ confidence limits for the cosmological parameters  using PLC+HST+WL+SZ data. 
The $\Delta \chi^2_{best}$ and the $\ln {B}_{ij}$ refer to the difference with respect to the $\Lambda$CDM.}
\label{tab:Tabel_PLC+HST+WL+SZ}}
\scalebox{0.9}{
\begin{tabular}{|c|c|c|c|}
\hline
\multicolumn{4}{|c|}{dataset: PLC+HST+WL+SZ}\\
\hline
{Parameter}&
{\textbf{$\Lambda$CDM }}&
{\textbf{$\Lambda$CDM + $N_{eff}$}}& 
{\textbf{HZP+$N_{eff}$}}
\\
\hline
$100\,\Omega_b h^2$ 	
& $2.251 \pm 0.021$ %
& $2.276 \pm 0.025$ %
& $2.295 \pm 0.021$ %
\\
$\Omega_{c} h^2$	
& $0.1160 \pm 0.0019$ %
& $0.1227 \pm 0.0037$ %
& $0.1229 \pm 0.0031$ 
\\
$100\, \theta$ 
& $1.04120 \pm 0.00043$
& $1.04064 \pm 0.00050$ 
& $1.04065 \pm 0.00052$ 
\\
$\tau$
& $0.082 \pm 0.015$
& $0.093 \pm 0.017$ 
& $0.099 \pm 0.018$
\\
$\ln 10^{10}A_s$  \footnotemark[1]
\footnotetext[1]{$k_0 = 0.05\,\Mpc^{-1}$.}
& $3.088 \pm 0.029$ 
& $3.128 \pm 0.035$ 
& $3.149 \pm 0.033$ 
\\
$n_{s}$
& $ 0.974 \pm 0.0051$ 
& $ 0.989 \pm 0.0080$ 
& $-$ 
\\
$N_{eff}$
& fixed to $3.046$
& $ 3.50 \pm 0.20$ 
& $ 3.67 \pm 0.13$ 
\\
$H_0$
& $ 68.99 \pm 0.79 $ 
& $ 71.90 \pm 1.54 $ 
& $ 73.63 \pm 0.61 $ 
\\
$\sigma_{8}$
& $ 0.817 \pm 0.012$
& $ 0.847 \pm 0.018$  
& $ 0.857 \pm 0.014$ 
\\
\hline
\hline
$\Delta \chi^2_{\rm best}$         
& $ - $	
& $ + 0.7 $ 
& $ + 7.7 $ 
\\
$\ln \mathit{B}_{ij}$ 
& $-$  
& $+ 2.6$  
& $+ 5.0$  
\\
\hline
\end{tabular}}
\end{table} 

\section{Method}
We consider two models, namely the HZP+$N_{eff}$ and the minimal $\Lambda$CDM, parameterised with the usual set of cosmological parameters: the baryon density, $\Omega_bh^2$, the cold dark matter density, $\Omega_ch^2$, the ratio between the sound horizon and the angular diameter distance at decoupling, $\theta$, the optical depth, $\tau$, the primordial scalar amplitude, $A_s$, and the primordial spectral index $n_s$. 
For the HZP+$N_{eff}$ model we consider $n_s$ fixed to unity and add the effective number of relativistic degrees of freedom as free parameter.
We choose to work with flat priors {{(see Tab.~\ref{tab:priors})}} and purely adiabatic initial conditions, also fixing the sum of neutrino masses to $0.06~eV$.

We perform our {{Bayesian}} analysis employing the nested sampling algorithm of MultiNest code~\cite{Feroz:2008xx,Feroz:2007kg,Feroz:2013hea} in the most recent release of the package CosmoMC~\cite{Lewis:2002ah}. {{We use the Boltzmann solver CAMB code~\cite{camb}, already implemented in CosmoMC, to compute the theoretical predictions of the models. The Bayesian evidence of our analysis are obtained using the Nested Importance Sampling (INS)~\cite{Feroz:2013hea, Cameron:2013sm} of MultiNest.}}

We consider the base dataset, ``PLC+HST", composed by the CMB data from the latest Planck Collaboration (2015) release~\cite{Aghanim:2015xee, Ade:2015xua} (PLC), and the prior on the local value of the expansion rate 
based on direct measurements made with the HST~\cite{Riess}. The PLC likelihood is composed of high-$\ell$ Planck temperature data and the low-P data by the joint TT,EE,BB and TE likelihood\footnote{ In addition to the usual cosmological parameters, we also vary the nuisance foregrounds parameters~\cite{Aghanim:2015xee}.}. 
We extend the base dataset with BAO 2PCF data obtained from the 6dF Galaxy Survey (6dFGS)~\cite{Beutler:2011hx}, SDSS DR7 Main Galaxy Sample (SDSS- MGS) galaxies~\cite{Ross:2014qpa},  BOSS galaxy samples, LOWZ and CMASS~\cite{Anderson:2013zyy}. 
The BAO 2PACF data we consider here are from recent analysis of $2$-point angular correlation function of luminous red galaxies from the SDSS-DR10 and DR11~\cite{gabimicol, gabi, newgabi}.
In order to improve the constraints on $\sigma_{8}$ we also include in our dataset  the thermal Sunyaev-Zeldovich (SZ) counts observations from Planck~\cite{Ade:2015fva}\footnote{We also consider the scaling relation parameters in our analysis, as extensively described in Ref.~\cite{salvati}.} and the low redshift weak gravitational lensing (WL) data from The Canada-France-Hawaii Telescope Lensing
Survey CFHTLenS, which is  a  $154$  square  degree  multi-filter  survey \cite{Heymans:2012gg} that provides one of the most robust low-redshift measures of cosmology to date.

{For the Bayesian model comparison we use the revisited version of the Jeffreys scale~\cite{Jeffreys} suggested in \cite{Trotta}, where the preference  of the analysed model with respect to the reference one is classified as: ${\textit{inconclusive}}$ for $\ln \mathit{B}_{ij}  = 0 - 1$,  ${\textit{weak}}$ for $ \ln \mathit{B}_{ij}  = 1 - 2.5 $,  ${\textit{moderate}}$  for $ \ln \mathit{B}_{ij}  = 2.5 - 5 $ and ${\textit{strong}}$ for $ \ln\mathit{B}_{ij}  > 5 $, 
being $\mathit{B}_{ij}$ the Bayes factor, i.e. the ratio between the Bayesian evidence of the analised model over the reference one. Note that a negative $\mathit{B}_{ij}$ means  preference of the latter over the former model. 

 \begin{figure}[!]
 \center
   \includegraphics[scale=0.7]{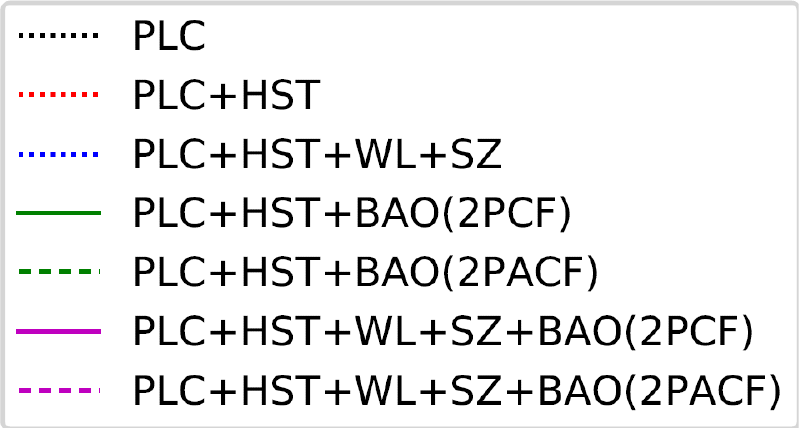}
 \includegraphics[scale=0.7]{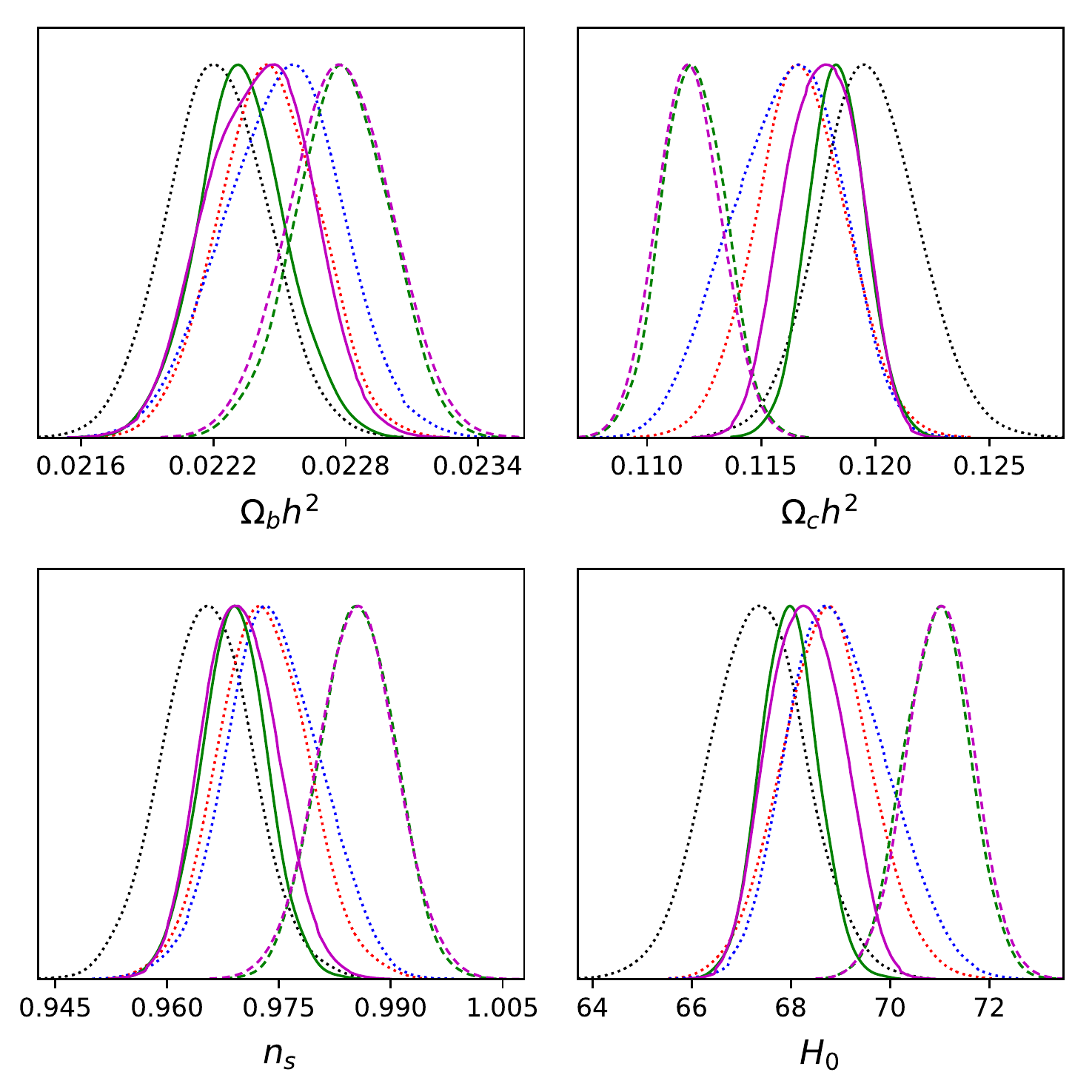}

 \caption{One-dimensional probability distribution for the cosmological parameters of the standard cosmological model, $\Lambda$CDM, using several dataset.}
\label{fig:LCDMs}
 \end{figure}
 
\section{Results}

The main quantitative results of our analysis are displayed in Table~\ref{tab:Tabel_PLC+HST(+BAO)}, \ref{tab:Tabel_PLC+HST+WL+SZ(+BAO)} and \ref{tab:Tabel_PLC+HST+WL+SZ}, where the values of $\Delta\chi^2$ and Bayes factor shown are calculated with respect to the minimal $\Lambda$CDM scenario using the same dataset. From Table~\ref{tab:Tabel_PLC+HST(+BAO)}, it is clear that for the $\Lambda$CDM model the inclusion of the BAO 2PCF data does not alter significantly the parameter estimation found in previous analysis of~\cite{Benetti:2017gvm, Ade:2015xua, Addison:2017fdm} while the dataset PLC+HST+BAO 2PACF imposes new constraints on the cosmological parameters (see also Fig.~\ref{fig:LCDMs}). In particular, it is worth mentioning that the lower the value of $\Omega_{c}h^{2}$  the higher the value of $n_{s}$. We also note that the BAO 2PACF data allows for a Hubble parameter value of $H_0 = 70.94 \pm 0.63$ $\rm{km.s^{-1}.Mpc^{-1}}$ (68\% C.L.), which alleviates the $H_0$ tension.
This difference in the results may be due to the differences between the techniques used for measuring the BAO signatures, as the BAO 2PACF data are obtained almost model-independently whereas the BAO 2PCF measurements adopt the $\Lambda$CDM model as the fiducial cosmology (we refer the reader to \cite{gabimicol}). 

Concerning the HZP+$N_{eff}$ model, both analyses using BAO 2PCF and BAO 2PACF point to a higher value of $\Omega_{c}h^{2}$, and the derived $H_0$ value fully agrees with the HST measurement \cite{Riess,Riess:2018byc}, at the cost of a higher value of $N_{eff}$. 
{{We also find that the BAO 2PCF measurements strongly disfavor the HZP+$N_{eff}$ model ($\ln B_{ij}=-7.6$), while the BAO 2PACF signature favors the latter in comparison to the $\Lambda$CDM scenario ($\ln B_{ij}=+6.3$).}} Such results suggest that the BAO 2PACF are in better agreement with the local measurements of the Hubble constant than the BAO 2PCF data, which in turn agrees with current CMB Planck (2015) constraints. We also calculate  the $\sigma_{8}$ best-fit value for the HZP+$N_{eff}$ model and note that it changes from $\sigma_{8} =0.85$ for the dataset with BAO 2PCF to $\sigma_{8} = 0.89$ for the dataset with BAO 2PACF. These two values are bigger than the marginalised ones obtained for the $\Lambda$CDM model using Planck data, i.e., $\sigma_{8} = 0.829 \pm 0.014$~\cite{Ade:2015xua}. 
This is probably due, as mentioned earlier, to the value of the {{Gaussian}} prior on the Hubble constant adopted in the analysis. 

We performed a second analysis adding the lensing data and Sunyaev-Zeldovich cluster counts. The results are summarized  in Table \ref{tab:Tabel_PLC+HST+WL+SZ(+BAO)}. The dataset PLC+HST+WL+SZ+BAO 2PACF provides different constraints on the cosmological parameter for the standard model with respect to the respective  2PCF data, which require a lower value for $\Omega_{c}h^{2}$ and a higher value of  $n_{s}$, as in the previous case.  In agreement with the results shown Table I, we note that the BAO 2PACF data allows values of the Hubble constant closer to the HST measurement, and  that for the HZP+$N_{eff}$ model,  $N_{eff}$ assumes significantly lower values using  BAO 2PACF dataset in comparison to the value obtained using BAO 2PCF. Such value matches the Planck Collaboration analysis at $2\sigma$, considering only the PLC data~\cite{Ade:2015xua}.
As mentioned before, although considerably far from the standard value,
$N_{eff} = 3.046$, recent BBN and CMB data also allow
for values of $N_{eff}> 3$ (see, e.g.~\cite{Ade:2015xua, DiValentino:2016hlg, Nollett:2013pwa,Benetti:2013wla, Hinshaw:2012aka, Boehm:2012gr, Cheng:2013csa}). 

Concerning the value of $\sigma_{8}$, we observe that the BAO 2PACF analysis prefers an higher value of $H_{0}$  and a lower value of $\sigma_{8}$, when compared to the one obtained with BAO 2PCF.  However by comparing the  $\Lambda$CDM model with the HZP+$N_{eff}$  we see that both  $H_{0}$ and $\sigma_{8}$ values are higher in the later.  {{The Bayesian analysis, by including WL+SZ dataset, essentially confirms our previous result: the HZP$+N_{eff}$ model is disfavored ($\ln B_{ij}=-3.9$) with respect to the $\Lambda$CDM when the BAO 2PCF are considered, while it becomes strongly favored using the BAO 2PACF ($\ln B_{ij}=+5.1$).}}

{{Finally, in order to better understand the tension between the dataset, we also perform an analysis using  only PLC+HST+WL+SZ. The results are reported in Tab.~\ref{tab:Tabel_PLC+HST+WL+SZ}. We note the agreement of the Bayesian preference for the HZP+$N_{eff}$ model with the analysis in which BAO (2PACF) data are considered, while the constraints on the parameters confirm the results obtained using PLC+HST data~\cite{Benetti:2017gvm}
(see also Fig.~\ref{fig:LCDMs}). These results highlight that all the different combinations of datasets used prefer the HZP+$N_{eff}$ over the $\Lambda$CDM, with the exception of the ones in which BAO (2PCF) data are considered. In other words, while the PLC and BAO (2PCF) support the $\Lambda$CDM model, the HST and BAO (2PACF) data suggest an upper value of $n_s$ and $H_0$. Interesting, the analysis of the $\Lambda$CDM+$N_{eff}$ model of Tab.~\ref{tab:Tabel_PLC+HST+WL+SZ}} shows a higher value for the spectral index  $n_s = 0.9901 \pm 0.0085$ and $H_0= 72.06 \pm 1.56$.

We emphasize that the significant difference in our results using BAO (2PCF) and (2PACF)  is due to the the deeply different measurements techniques of BAO signal. In particular, while the BAO (2PCF) use a fiducial model to estimate the cosmological distance (i.e. without error bars) between the galaxies, the BAO (2PACF) use the angle separation information, with its associated error. Also, the BAO (2PCF) use information of parallel and transversal modes to estimate the measurements, while the BAO (2PACF) use only the transversal model (being more model independent). Finally, the BAO (2PCF) use a deep range of redshifts to perform the measurements, while the BAO (2PACF) use very narrow redshift shells, so that its depth can be  considered as being $\delta z \sim 0$. 
This leads to error bars of one order of magnitude bigger in the BAO (2PACF) case, and also differences in the mean value since all the information of 11 BAO measurements of BAO (2PACF) in the range $0.45 < z < 0.65$ are compressed in the only  measurement of CMASS BAO (2PCF), calculated in the range $0.43<z<0.70$. {{From a simple average of the values of the 11 BAO (2PACF) measurements, we found $D_A (0.45 \leq z \leq 0.65)/r_s = 8.47 \pm 0.67$, which agrees only at 2$\sigma$ with the CMASS mean value $D_A(z = 0.57)/r_s = 9.524 \pm 0.015$.}} 

\section{Conclusions}

In this work we analysed observational constraints on the standard $\Lambda$CDM scenario and an exact scale invariant scenario in which the number of effective degrees of freedom is taken as a free parameter, HZP$+N_{eff}$. Although a exact scale invariant spectrum lacks a theoretical justification,  ruling out such a scenario as well as its extensions is important to confirm the expectation of small deviations from scale invariance, which are generic to all inflationary models.  We have extended the results of Ref.~\cite{Benetti:2017gvm} taking into account both the $H_0$ and the $\sigma_8$ tensions and examined in detail the viability of the HZP+$N_{eff}$ model using the latest data sets of CMB, BAO, WL, GC and $H_0$. 

First, using as base dataset the joined CMB and HST data, we explored the role of 2PCF and 2PACF BAO measurements on the cosmological analysis. {{We obtained a higher value of Bayesian evidence for the minimal $\Lambda$CDM scenario using the BAO 2PCF measurements compared to the value obtained using the BAO 2PACF data}}. Also, while the $\Lambda$CDM model is preferred by the BAO 2PCF, the HZP+$N_{eff}$ model becomes favored when the  BAO 2PACF data are considered, which is in agreement with our previous results using only PCL+HST data. Both BAO data sets favor an higher $H_{0}$ value, which alleviates the tension between the Planck 2015 CMB data and the HST value of $H_0$.  Second, including into the previous dataset  cluster counts and weak lensing observations, in order to better constrain the $\sigma_8$ parameter, the new analysis basically confirms the previous result.
{{Finally, we performed an analysis  using only PLC+HST+WL+SZ data and found its results in agreement with the ones using BAO (2PACF). This analysis helps better understand the differences in the results: the PLC and BAO (2PCF) support the $\Lambda$CDM model while the HST and 2PACF prefer an upper value for both $n_s$ and $H_0$. We emphasize that this is an important aspect of the present analysis since both HST and 2PACF BAO measurements are model-independent and the latest results from 50 Cepheids using Gaia DR2 parallaxes and HST photometry not only confirm the $H_0$ tension but also enhance it (3.8$\sigma$)~\cite{Riess:2018byc}.}}   Our analysis, therefore, provides a clear example of a theoretically unmotivated model which can provide a good fit to current observations given the current tensions between different data sets and measurements. Such results clearly  reinforces the
need for more precise, accurate and model-independent
observations of the large-scale structure and local expansion rate than currently available.

{\it{Acknowledgments}} --
The authors  thank Gabriela C. Carvalho for useful discussions.
MB and LG acknowledges financial support of the Funda\c{c}\~{a}o Carlos Chagas Filho de Amparo \`{a} Pesquisa do Estado do Rio de Janeiro (FAPERJ - fellowship {\textit{Nota 10}}). 
JSA acknowledges support from the Conselho Nacional de Desenvolvimento Cient\'{i}fico e Tecnol\'{o}gico (CNPq) (Grants no. 310790/2014-0 and 400471/2014-0) and FAPERJ (grant no. 204282). 
We also acknowledge the authors of the CosmoMC (A. Lewis) and Multinest (F. Feroz) codes. 

\end{document}